# Patterns of Selection of Human Movements IV: Energy Efficiency, Mechanical Advantage, and Asynchronous Arm-Cranking

**Stuart Hagler**

Oregon Health & Science University

Portland, OR, USA

haglers@ohsu.edu

**Abstract:** Human movements are physical processes combining the classical mechanics of the human body moving in space and the biomechanics of the muscles generating the forces acting on the body under sophisticated sensory-motor control. The characterization of the performance of human movements is a problem with important applications in clinical and sports research. One way to characterize movement performance is through measures of energy efficiency that relate the mechanical energy of the body and metabolic energy expended by the muscles. Such a characterization provides information about the performance of a movement insofar as subjects select movements with the aim of maximizing the energy efficiency. We examine the case of the energy efficiency of asynchronous arm-cranking doing external mechanical work, that is, using the arms to turn an asynchronous arm-crank that performs external mechanical work. We construct a metabolic energy model and use it estimate how cranking may be performed with maximum energy efficiency and recover the intuitive result that for larger external forces the crank-handles should be placed as far from the center of the crank as is comfortable for the subject to turn. We further examine mechanical advantage in asynchronous arm-cranking by constructing an idealized system that is driven by a crank and which involves an adjustable mechanical advantage and analyze the case in which the avg. frequency is fixed and derive the mechanical advantages that maximize energy efficiency.

## 1 Introduction

Human movements are physical processes combining the classical mechanics of the human body moving through space with the biomechanics of the muscles generating the required forces under the control of sophisticated, coordinated sensory-motor and cognitive processes. [1, 2] Due to the sophisticated nature of human movements, it can be difficult to characterize the quality of their performance, though concepts such as "efficiency," "economy," and "effectiveness" have been proposed. [3] One approach to characterizing performance is to ignore the sensory-motor and cognitive aspects of motor control, and focus on the physical mechanics of movement by using one or more measures of energy efficiency that relate the mechanical energy of the body moving through space with the metabolic energy expended by the muscles to generate the forces. In the previous papers in this series, [4-6] we analyzed human walking gait focusing in [6] on the energy efficiency of walking gait. In this paper, we complement the examination in [6] of the energy efficiency of walking gait by looking at the energy efficiency of a different movement – asynchronous arm-cranking. The goal of this project is to examine how maximizing energy efficiency shapes the patterns of human movements using the case of asynchronous arm-cranking.

An arm-crank is a mechanical device that a subject can grab hold of by using the hands and can turn by moving the arms so that it does external mechanical work. Arm-cranks take two forms – the asynchronous and synchronous arm-crank. [7-10] In the asynchronous arm-crank the two crank-handles which the hands grab onto are attached to two crank-arms that extend in opposite directions from the center of the crank and around which the crank rotates (this resembles the leg-crank pedaling system that is used to power a typical bicycle). In the synchronous arm-crank, the two crank-handles attach to either side of a single crank-arm that extends from the center of the crank and around which the crank rotates. While any number of mechanical devices may be built around either type of arm-crank, two widespread applications are: (i) in exercise or rehabilitation in the form of the upper body ergometer which works the arms and may be used as an alternative to treadmill exercise for clinical evaluation, [11, 12] and rehabilitation, [13] and (ii) in sports or recreation in the form of handcycling involving any number of simple bicycle-like vehicles powered using an arm-crank. [14]

In this paper, we complement the project in [6] of looking at the patterns of movement that maximize energy efficiency in walking gait by examining the patterns of movement in asynchronous arm-cranking that do the same. The approach we continue to take is to use models of the metabolic energy of movements doing external mechanical work to construct models of the energy efficiency from which we derive the movement patterns that maximize the energy efficiency. We carry out this project out in four parts. In the first part (Sec. 2), we provide an outline of the formalism we use to describe the metabolic energies and external mechanical work, reviewing the relevant models developed in [4, 5] and defining four measures of energy efficiency following Gaesser & Brooks [15] that we developed in [6]: gross energy, net energy, work, and delta work efficiency. In the second part (Sec. 3), we develop a model of the metabolic energies of asynchronous arm-cranking using the metabolic energy formalism and plug this model into the movement utility formalism to construct a model to describe the selection of cranking. We validate this model by showing that the metabolic energy model can be fit well to the available empirical data in Atzler et al. [16] and use this fit to provide empirical values to free parameters in the model. In the third part (Sec. 4), we use the model to estimate the four measures of energy efficiency as functions of the external force for asynchronous arm-cranking doing external mechanical work. We find that the estimated work and delta work efficiencies agree with estimates in Powers et al. [17] We then derive the crank-arm radii that maximize the net energy efficiency given a fixed avg. frequency of rotation of the crank. In the fourth part (Sec. 5), we modify the situation in Sec. 4 by introducing an idealized adjustable mechanical advantage that is similar to an idealized device developed in [6].

## 2 Energy Efficiencies of a Movement

We begin with the first part of this project in which we develop models of the metabolic energy and external mechanical work of a movement and relate them using four measures of energy efficiency. We first review the approach to modeling human movements that we developed in [4, 5] including external mechanical work as developed in [5]. We then define four measures of energy efficiency of human movements following Gaesser & Brooks [15] that allow us to relate the metabolic energy of a movement to the external mechanical work done by the movement.

*2.1 Metabolic Energy*

We model the human body executing a movement as in [4, 5] using a skeleton consisting of a system of $\mathrm{N}$ ("nu") segments attached at $N$ joints. A movement causes the joints to move with joint-trajectories so that the *n-th* joint moves with joint-trajectory $\vec{x}_n(t)$. The metabolic rate $\dot{W}(t)$ of expending metabolic energy is the sum of metabolic rates $\dot{W}_n(t)$ associated with each joint $\dot{W}(t) \approx \sum_{n=1}^{N} \dot{W}_n(t)$. The metabolic rate associated with a joint is given by a function:

$$\dot{W}_n(t) = \dot{W}_n^F\left(\vec{F}_n(t)\right) + \dot{W}_n^E\left(\vec{F}_n(t), \vec{v}_n(t)\right). \tag{1}$$

The metabolic rates $\dot{W}_n^F(t)$ and $\dot{W}_n^E(t)$ have the mathematical forms:

$$\begin{aligned}\dot{W}_n^F\left(\vec{F}_n(t)\right) &\approx \varepsilon_n F_n(t)^2, \\ \dot{W}_n^E\left(\vec{F}_n(t), \vec{v}_n(t)\right) &\approx \eta_n \vec{F}_n(t) \cdot \vec{v}_n(t).\end{aligned} \tag{2}$$

The quantities $\varepsilon_n$ and $\eta_n$ are constant parameter values characterizing the associated metabolic rates. The parameters $\eta_n$ may take on different values when the muscles add or remove mechanical energy to or from a segment, though we require they be constant in each case. The mechanical energy is *positive* when mechanical energy is added to a segment and *negative* when it is removed.

*2.2 External Mechanical Work*

The *external mechanical work* $U_{ext}(t)$ is mechanical energy transferred from the segments of the body to objects external to the body with the intent of carrying out a task using those objects. The external mechanical work is *positive* when mechanical energy is added to an object and *negative* when it is removed.

*2.3 Measures of Energy Efficiency*

We define four measures of the energy efficiency following Gaesser & Brooks: [15] (i) *gross energy efficiency* $\upsilon_{gross}$ (external mechanical work done divided by gross metabolic energy expended), (ii) *net energy efficiency* $\upsilon_{net}$ (external mechanical work done divided by metabolic energy expended), (iii) *work efficiency* $\upsilon_{work}$ (external mechanical work done divided by metabolic energy expended above the case when no external mechanical work is done), (iv) *delta work efficiency* $\upsilon_{delta}$ (change in external mechanical work done divided by change metabolic energy as external mechanical work rate changes). For a subject having a resting metabolic rate $\dot{W}_{rest}$ and a movement doing external mechanical work $U_{ext}$ in a time $T$, and expending metabolic energy of $W(U_{ext})$, the four measures of energy efficiency are:

$$\begin{aligned}\upsilon_{gross}(U_{ext}) &= U_{ext} / \left(\dot{W}_{rest} T + W(U_{ext})\right), \\ \upsilon_{net}(U_{ext}) &= U_{ext} / W(U_{ext}), \\ \upsilon_{work}(U_{ext}) &= U_{ext} / \left(W(U_{ext}) - W(0)\right), \\ \upsilon_{delta}(U_{ext}) &= \left(\frac{dW}{dU_{ext}}(U_{ext})\right)^{-1}.\end{aligned} \tag{3}$$

**3 Asynchronous Arm-Cranking Doing External Work**

In the second part of the project, we develop models for the metabolic and mechanical energies in asynchronous arm-cranking for the case where external work is done. We proceed in the following stages: (i) we give a kinematic model of asynchronous arm-cranking (Sec. 3.2), (ii) we construct a metabolic energy model for asynchronous arm-cranking (Sec. 3.3), (iii) we discuss how the metabolic energy model for asynchronous arm-cranking that we have constructed approximates the physical situation of cranking and what additional mathematical terms might be required to construct a more complete model (Sec. 3.4), and (iv) we use available empirical data in Atzler et al. [16] to estimate values for the free parameters in the metabolic energy model (Sec. 3.5).

*3.1 Some Anthropometric Values*

For convenience, we give here, in one place, several relevant anthropometric values. A subject has mass $M$ and height $H$. Using typical anthropometric values, this means that the subject's waist is a height $\rho_{waist}H$ from the floor where $\rho_{waist} \approx 0.53$ the subject's shoulders are a height $\rho_{shoulder}H$ from the floor where $\rho_{shoulder} \approx 0.82$ from the floor, and the subject's arms are a length $\rho_{arm}H$ where $\rho_{arm} \approx 0.33$. [18] The mass in each arm (i.e. upper arm, lower arm, and hand) of about $m = 0.05M$. [18]

*3.2 Kinematic Model*

The subject turns an asynchronous arm-crank with the crank-handles placed at the ends of crank-arms that extend a radius $R$ from the center; the center is placed at a height $H_{crank}$ from the floor. We normalize the height $H_{crank}$ of the crank in terms of the subject's height $H$ using $\rho = H_{crank}/H$. The crank rotates so that the hands move in planes that are parallel to the sagittal plane of the body. For convenience, we describe asynchronous arm-cranking as occurring in the two dimensions of the sagittal plane. We use a three-segment model with one segment for each arm and one segment for the body (i.e. legs, torso, and head). The arms are joined to the body at a point corresponding to the shoulders. The subject is free to adjust the body to place the shoulders at any physically reasonable position relative to the crank. This may be done using any combination of bending the knees, spreading the legs apart or bending the body at the hips (see [16] for pictures of a subject orients the body in practice). The body does not move during cranking so we indicate the body by a single segment placed so that the shoulders are in the required position. The arms attach to the crank-handles of the crank, and we denote the point where they attach as the *hands*. The model is two dimensional, so we arbitrarily label one of the hands as the *right hand* and the other as the *left hand*. The arms are straight lines between the hands lying on the crank-handles of the crank and the shoulders but may change in length during the movement. The maximum length each arm takes should correspond to the physical length of the arms while the shortest length should correspond to a physically reasonable bending of the arm at the elbow. The subject turns the crank by using the arms to rotate the crank-handles around the center at a frequency $f$.

We only look at steady state asynchronous arm-cranking that is in progress and maintains constant values for the movement parameters; we do not look at the process of starting or stopping asynchronous arm-cranking. We describe cranking using two movement parameters: (i) the avg. frequency $f$ at which the asynchronous arm-crank turns, and (ii) the radius $R$ from the center of the crank of the crank-handles on the crank-arms. We define the sagittal plane using the unit vectors $\hat{x}$ and $\hat{z}$ indicating the horizontal and vertical directions, respectively. We use the following model to describe the motions of the right hand and left hand:

$$\begin{aligned}
\vec{x}_r(t) &= R\cos(2\pi ft + \pi/2)\hat{x} + R\sin(2\pi ft + \pi/2)\hat{z}, \\
\vec{x}_l(t) &= R\cos(2\pi ft + 3\pi/2)\hat{x} + R\sin(2\pi ft + 3\pi/2)\hat{z}.
\end{aligned} \quad (4)$$

This places the center of the crank at the origin with the vectors pointing from the center of the crank to the crank-handles given by (4). For convenience, and without a loss of generality, we have defined time zero in (4) to lie at a particular point in the cranking cycle.

*3.3 Metabolic Energy Model*

We place all the mass $m$ of each arm in the corresponding hand. We assume that the asynchronous arm-crank provides the force needed to hold up the hands against gravity and that the mechanical energy is conserved so that any mechanical energy that needs to be added to the hand moving upward against gravity comes from the corresponding mechanical energy gained by the hand moving downward with gravity. We denote the force applied by muscles to the right hand and thus to the asynchronous arm-crank by $\vec{F}_r(t)$ and that applied to the left hand and thus to the asynchronous arm-crank by $\vec{F}_l(t)$. We assume the hands and arms are identical, and associate parameters $\varepsilon_h$ and $\eta_h$ with each hand; these parameters correspond to the parameters in (2). We define the external force $F_{ext}$ so that it generates a torque that opposes the rotation of the crank, and assume that each hand generates a continuous force $F_{ext}/2$ in the direction of the rotation to compensate for the external force; that is, the forces generated by each hand to oppose the external force are perpendicular to the vectors defined in (4). The time required to complete a turn is $T = 1/f$. The metabolic energy per turn is the sum of a constant term $W_0$, and four metabolic energies: (i) the energy expended generating the force $\vec{F}_r(t)$ of the right hand, (ii) the energy expended generating the force $\vec{F}_l(t)$ of the left hand, (iii) the energy expended by the right hand to provide the mechanical energy of the external mechanical work, and (iv) the energy expended by the left hand to provide the mechanical energy of the external mechanical work; this is:

$$W(f, R, F_{ext}) \approx W_0 + \varepsilon_h \int_0^{1/f} F_r(t)^2 \, dt + \varepsilon_h \int_0^{1/f} F_l(t)^2 \, dt \\ + \eta_h \cdot (F_{ext}/2)(2\pi R f) \int_0^{1/f} dt + \eta_h \cdot (F_{ext}/2)(2\pi R f) \int_0^{1/f} dt. \quad (5)$$

In addition to the force $F_{ext}/2$ in the direction of the rotation of the asynchronous arm-crank, each hand must also generate a force acting toward the center of the crank so that the hands revolve along a circle around the center; we call this force the *centripetal force* $F_{cent}(f, R)$, and it acts parallel to the vectors defined in (4). We illustrate this in Fig. 1. The instantaneous speed of each hand is $v = 2\pi R f$. For each hand to revolve smoothly in a circle around the center of the crank, the centripetal force must satisfy $F_{cent}(f, R) = mv^2/R$, that is:

$$F_{cent}(f, R) = 4\pi^2 m R f^2. \quad (6)$$

For each hand the force $F_{ext}/2$ in the direction of the rotation of the crank is perpendicular to the centripetal force $F_{cent}(f, R)$, therefore the squares of the forces $\vec{F}_r(t)$ and $\vec{F}_l(t)$ are given by:

$$F_r(f, R, F_{ext})^2 = F_{cent}(f, R)^2 + F_{ext}^2/4, \\ F_l(f, R, F_{ext})^2 = F_{cent}(f, R)^2 + F_{ext}^2/4. \quad (7)$$

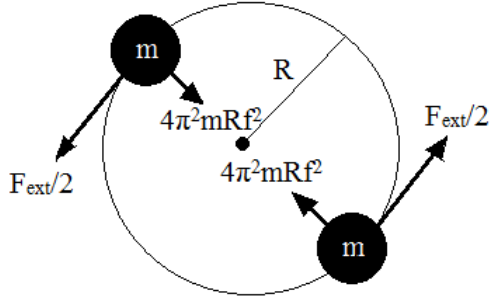

**Figure 1.** The motion of the hands around the center of the asynchronous arm-crank. The crank rotates at a constant frequency $f$. To maintain the constant rotational motion, the muscles must provide a centripetal force to carry the hands around the center of the crank at the constant radius $R$ of the crank-handles on the crank-arms.

Combining (5), (6), and (7), we find that the metabolic energy per turn satisfies:

$$W(f, R, F_{ext}) \approx W_0 + \left(32\pi^4 \varepsilon_h m^2\right) R^2 f^3 \\ + \left(\varepsilon_h / 2\right) F_{ext}^2 / f + \left(2\pi \eta_h\right) F_{ext} R. \tag{8}$$

We find it adds to conceptual clarity to define the two quantities:

$$\begin{aligned} W_{turn}(f, R) &= W(f, R, 0), \\ W_{ext}(f, R, F_{ext}) &= W(f, R, F_{ext}) - W(f, R, 0). \end{aligned} \tag{9}$$

In this way, $W_{turn}(f, R)$ gives the metabolic energy per turn to turn the asynchronous arm-crank without doing any external mechanical work, while $W_{ext}(f, R, F_{ext})$ gives the metabolic energy per turn beyond $W_{turn}(f, R)$ needed to generate the external force $F_{ext}$. We can use (9) to write (8) compactly as:

$$W(f, R, F_{ext}) \approx W_{turn}(f, R) + W_{ext}(f, R, F_{ext}). \tag{10}$$

*3.4 Corrected Metabolic Energy Model*

While we expect that the metabolic energy model in (10) accounts for much of the metabolic energy of asynchronous arm-cranking, we also expect that there are further motions of the body and muscle forces that the model has ignored. We can complete the description formally by introducing an unknown function $W_{ext}^{(1)}(f, R, F_{ext})$ that is required to satisfy $W_{ext}^{(1)}(f, R, 0) = 0$ so that the uncorrected metabolic energy model in (10) is replaced with a corrected metabolic energy model of the form:

$$W(f, R, F_{ext}) = W_{turn}(f, R) + W_{ext}(f, R, F_{ext}) \\ + W_{ext}^{(1)}(f, R, F_{ext}). \tag{11}$$

Ideally, since we do not know the form of $W_{ext}^{(1)}(f, R, F_{ext})$, we would like its contribution to the corrected metabolic energy model to be small so that (11) is approximately equal to (10). We have introduced the formal term $W_{ext}^{(1)}(f, R, F_{ext})$ largely for convenience of expression when discussing the performance of the model in Secs. 3.5 and 3.6.

*3.5 Empirical Study (Atzler et al., 1923)*

We now make the metabolic energy model developed in (8) into an estimator of the metabolic energy during asynchronous arm-cranking over the range of allowed cases by using empirical data to produce estimates for the values of the parameters $W_0$, $\varepsilon_h$, and $\eta_h$.

Atzler et al. [16] looked at one subject (male: $M = 60$ kg, $H = 1.7$ m, age = 35 years, resting metabolic rate of $\dot{w}_{rest} = \dot{W}_{rest}/M = 0.30$ cal·kg⁻¹·s⁻¹). This subject is the same subject analyzed in [4-6, 19] although Atzler et al.'s [16] study took place about four years earlier than that in [19]. Metabolic energies were determined from expired gasses using a Benedict respiratory apparatus. The subject was trained to turn an asynchronous arm-crank at a frequency of $f = 18$ rpm (0.30 s⁻¹) at all vertices of a grid of 4 heights $H_{crank}$ (0.55 m, 0.83 m, 1.1 m, and 1.6 m corresponding to $\rho = H_{crank}/H$ of 0.33, 0.49, 0.67, and 0.95), 3 radii $R$ (0.19 m, 0.28 m, and 0.36 m), and performing 5 external works $U_{ext}$ per turn (64 J/turn, 130 J/turn, 190 J/turn, 250 J/turn, and 320 J/turn) giving 60 distinct cases. However, data were not collected for external works of 320 J/turn at heights of 1.1 m and 1.6 m (cf. [19] where there were also cases the subject could not sustain for measurement) leaving 58 distinct cases that were recorded.

Rewriting the metabolic energy model in (8), we find:

$$W \approx W_0 + \left(\left(32\pi^4 m^2\right) R^2 f^3 + F_{ext}^2 / 2f\right)\varepsilon_h + \left(2\pi F_{ext} R\right)\eta_h. \tag{12}$$

In the study in [16], the asynchronous arm-crank was set up so that the subject had to perform an external mechanical work of $U_{ext}$ for each turn of the crank. In this setup, the external force $F_{ext}$ that must be generated to give the required external mechanical work $U_{ext}$ per turn is:

$$F_{ext} = U_{ext} / 2\pi R. \tag{13}$$

We can therefore rewrite the metabolic energy expended per turn in (12) in terms of the external mechanical work $U_{ext}$ per turn as:

$$W = W_0 + \left(32\pi^4 m^2 R^2 f^3 + \left(\frac{U_{ext}^2}{8\pi^2 R^2 f}\right)\right)\varepsilon_h + \left(U_{ext}\right)\eta_h. \tag{14}$$

We would like to use (14) to estimate the parameter values $W_0$, $\varepsilon_h$, and $\eta_h$ using all of the asynchronous arm-cranking data available. However, the results of doing this were not as good as we would have liked and appeared to indicate a dependence on the height $H_{crank}$ of the crank. Neither $H_{crank}$ nor $\rho$ appear in the metabolic energy model in (8), but one way to include them is to suppose the parameters $W_0$, $\varepsilon_h$, and $\eta_h$ are functions $W_0(\rho)$, $\varepsilon_h(\rho)$, and $\eta_h(\rho)$; this is the approach we adopt. We found that the simplest set of functions $W_0(\rho)$, $\varepsilon_h(\rho)$, and $\eta_h(\rho)$ that performed well the data was that where $\varepsilon_h(\rho)$ takes on two values – one for $\varepsilon_h(0.33 \leq \rho \leq 0.67$, i.e. below the shoulder height of $\rho_{shoulder} \approx 0.82)$ and one for $\varepsilon_h(\rho =$ 0.95, i. e. above shoulder height of $\rho_{shoulder} \approx 0.82$) — while $W_0$ and $\eta_h$ are constant for all crank heights; we discuss this further in Sec. 3.6.

We estimate values for $W_0$, $\varepsilon_h$, and $\eta_h$ using least squares regression to fit a single linear model derived from (14) that combines the data for $0.33 \leq \rho \leq 0.67$, and $\rho = 0.95$ using parameters $W_0$, $\varepsilon_h(0.33 \leq \rho \leq 0.67)$, $\varepsilon_h(\rho = 0.95)$, and $\eta_h$. Using ordinary least-squares regression, this model fit the data for Atzler et al.'s subject with $R^2 = 0.92$ and $p < 0.0001$, and the parameter values were:

$$\begin{aligned}
W_0 &\approx 12\ cal, \\
\varepsilon_h\left(0.33 \leq \rho \leq 0.67\right) &\approx 3.0 \times 10^{-3}\ cal \cdot N^{-2} \cdot s^{-1}, \\
\varepsilon_h\left(\rho = 0.95\right) &\approx 8.1 \times 10^{-3}\ cal \cdot N^{-2} \cdot s^{-1}, \\
\eta_h &\approx 1.2\ cal \cdot J^{-1}.
\end{aligned} \quad (15)$$

Inspection of the 95% confidence intervals showed that all parameters but the $W_0$ parameter were statistically significant. This suggests that the value of $W_0$ could be near zero and thus consistent with the model as we developed it without the $W_0$ term; we discuss this further in Sec. 3.6. The fit is illustrated in Fig. 2 using all the cases observed by Atzler et al. The estimated parameter values in (15) agree in order of magnitude with the values of $W_0 \approx 7.0$ cal, $\varepsilon_{st} \approx 1.9 \times 10^{-3}$ cal $\cdot N^{-2} \cdot s^{-1}$, $\varepsilon_{sw} \approx 2.6 \times 10^{-3}$ cal $\cdot N^{-2} \cdot s^{-1}$, and $\eta_{st} \approx 0.62$ cal $\cdot J^{-1}$ estimated in [5] using the data in [19]; we discuss this further in Sec. 3.6.

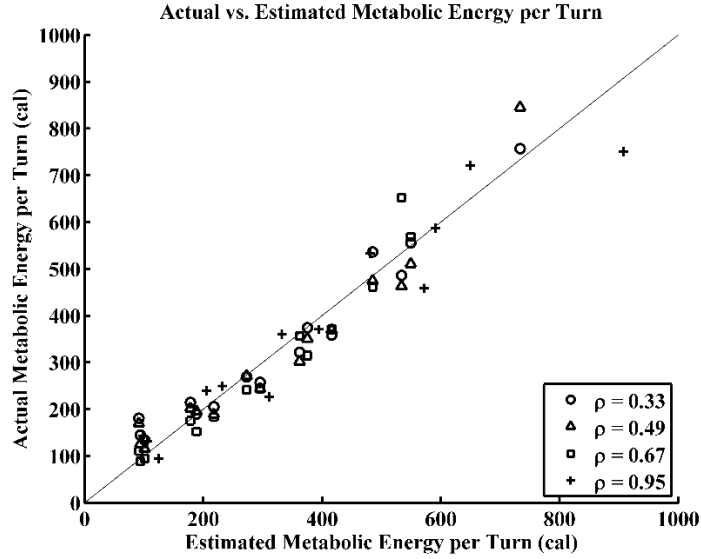

**Figure 2.** Actual vs. estimated metabolic energy per turn. We use the model in (14) to fit the observed metabolic energies for the 60 cases in Atzler et al. with ratio of the height of the asynchronous arm-crank from the floor to the subject's height $\rho = 0.33, 0.49, 0.67$, and $0.95$; the fit for the model for these walking gaits has $R^2 = 0.93$ and $p < 0.0001$. For reference, we show a segment of the line with slope one passing through the origin.

We may attribute the systematic deviations of the estimated metabolic energy per step from the actual metabolic energy per step in Fig. 2 to the absence of the function $W_{ext}^{(1)}(f, R, F_{ext})$ in the uncorrected metabolic energy model. To obtain some suggestion of the functional form of $W_{ext}^{(1)}(f, R, F_{ext})$ we can at these deviations as functions of the radius $R$, and the external force $F_{ext}$; we do this in Fig. 3.

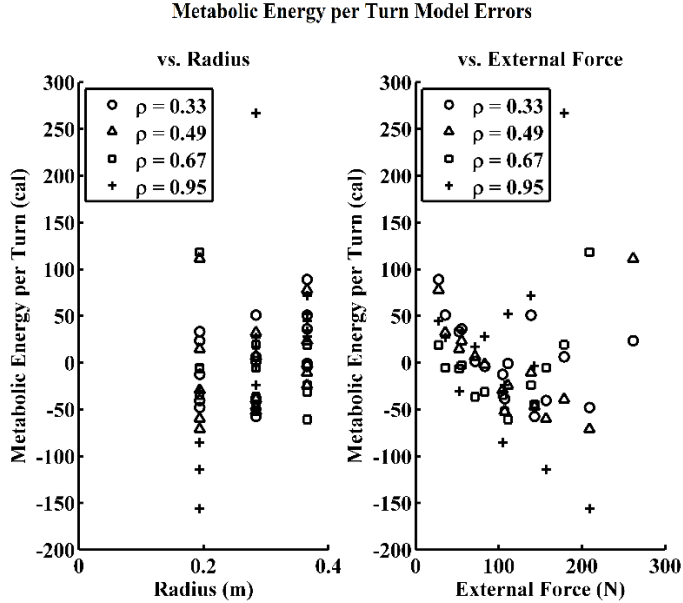

**Figure 3.** Metabolic energy per turn model errors. We suggest the functional form of $\boldsymbol{W_{ext}^{(1)}(f, R, F_{ext})}$ by plotting the differences in the actual metabolic energy per step and the metabolic energy per step as a function of the radius and the external force.

*3.6 Discussion*

We have assumed a relatively nice form for the forces of the arms on the asynchronous arm-crank to arrive at (7). The model of asynchronous arm-cranking that we have developed is the idealized case that requires that each hand maintains a constant magnitude force on the crank of constant which changes direction exactly as the crank rotates. It is possible that the forces the arms generate are not so nice so that either (i) the magnitudes of the forces on the crank vary over the course of the motion, or (ii) the arms generate additional forces that do not contribute to the asynchronous arm-cranking motion, but nevertheless require the expenditure of metabolic energy. We expect these sorts of effects to cause the model to produce estimates of the value of the parameter $\varepsilon_h$ in (15) that are somewhat higher than the true value.

We have approximated the distribution of mass in the arms by placing the entire mass of each arm in the hand. As the effect of more accurately modeling the distribution of mass in the arm would be to introduce a constant factor into the model whose only effect would be to change the estimated parameter values in (15), we have made this approximation of the distribution of mass in the body to arrive at a practical metabolic energy model for asynchronous arm-cranking, albeit one in which the estimated parameter values in (15) may be somewhat different from the true values. If we want to compare the estimates in (15) to the comparable parameter estimates for walking gait in [4-6] as we did in Sec. 3.5, we should model the distribution of mass in the body more accurately than we have (both here and in [4-6]).

We would have liked the model we used to estimate (15) to have constant values for the parameters $W_0$, $\varepsilon_h$, and $\eta_n$ that did not depend on the position of the asynchronous arm-crank. This would have reflected the model we developed in Sec. 3.3 being an approximately complete description of asynchronous arm-cranking. Instead we used a model in which the parameter $\varepsilon_h$ took on different values for the case where the crank was placed below or above shoulder-height for the subject while the parameters $W_0$, and $\eta_n$ were held constant for all placements of the crank. This allowed us to analyze all the cranking data in a single model and obtain reasonably good results. We may account for the change in value of the parameter $\varepsilon_h$ for different heights of the crank as due to either (i) a difference in how the arm muscles are employed holding up and moving the arm in the two cases, or (ii) the effects of variations in the forces applied by the hands to the crank over the course of a turn which is both different from the forces in the model we have developed in Sec. 3.3 and which differ from each other when the placement of the crank changes.

The value estimated for the $W_0$ parameter in (15) was not statistically significant so it could possibly have a negligible or zero value. We argued in [5] that the analogous parameter in the walking gait model measures the mechanical energy loss during walking. While there seems to be less opportunity for a mechanical energy loss in asynchronous arm-cranking (unless the asynchronous arm-crank is poorly constructed) an alternative explanation for a non-zero value would be that this represents the metabolic cost of muscle forces placed on the crank that go beyond what is needed to turn it in the ideal case that we have described here that assumes the subject generates the minimal forces necessary to get the hands to turn the crank. It is possible for a subject to apply larger forces to turn the crank and rely on the rigidity of the crank to allow these forces to turn the crank as intended.

We have estimated a parameter value for $\eta_n$ in (15) that is about twice the comparable parameter values estimated in [5]. As the parameter $\varepsilon_h$ accounts for the metabolic energy of producing the muscle forces we do not expect the parameter $\eta_h$ to depend on the specifics of the movement being executed, but rather expect it to reflect the maximum efficiency with which a group of muscles can produce mechanical energy. If we believe there are effects in the corrected metabolic energy model that become more significant with larger external forces $F_{ext}$, then the value of $\eta_h$ should be more easily measured in the limit of small external forces $F_{ext}$. The data obtained by Atzler et al. in [16] ranged from 28 N to 260 N. In [6], we encountered an issue in the analysis of walking gait where parameter values estimated in the case where external work was done produced poor values in the limit where no external work was done, and a similar effect may be occurring here. Following [6], we may try to account for the large estimated value of $\eta_n$ by supposing there is some correlation between the $W_{ext}^{(1)}(f, R, F_{ext})$ and $F_{ext}R$ so that the estimate of $\eta_h$ in (15) for external forces $F_{ext}$ the range from 28 N to 260 N overfit the data and fail to produce the appropriate limit as the external force goes to zero. Alternatively we my look at the derivation of the formulas in (2) from [4]. We derived the formulas in (2) for calculating metabolic work by approximating $\dot{W}_n^F\left(\vec{F}_n(t)\right)$ and $\dot{W}_n^E\left(\vec{F}_n(t), \vec{v}_n(t)\right)$ using Taylor series truncated to the lowest order non-zero term. It is possible that the correct description of asynchronous am-cranking requires the presence of higher order terms in the expansion or the presence of terms that are functions of variables other than the $\vec{v}_n(t)$ or $\vec{F}_n(t)$.

## 4 Energy Efficiency of Asynchronous Arm-Cranking

In the third part of this project, we use the metabolic energy and external mechanical work models to estimate the energy efficiency of asynchronous arm-cranking. We first calculate the functional forms of the gross energy, net energy, work, and delta work efficiencies for cranking doing external mechanical work, and compare the resulting energy efficiency estimates to values reported by Powers et al. [17] We then estimate the radius $R$ of the crank-handles on the crank-arms as a function of the external force $F_{ext}$ for a subject selecting cranking doing external mechanical work of maximum net energy efficiency.

*4.1 Measures of Energy Efficiency*

We now calculate the gross energy, net energy, work, and delta work efficiencies of asynchronous arm-cranking doing external mechanical work for cranking by the subject observed by Atzler et al. [16] using the definitions in (3). The metabolic energy $W(f, R, U_{ext})$ per turn as a function of the external mechanical work $U_{ext}$ per turn is:

$$W\left(f, R, U_{ext}\right) = W_0 + \left(32\pi^4 \varepsilon_h m^2\right) R^2 f^3 \\ + \left(\varepsilon_h / 8\pi^2\right) U_{ext}^2 / R^2 f + \left(\eta_h\right) U_{ext}. \tag{16}$$

The energy efficiencies are therefore:

$$\begin{aligned}
\upsilon_{gross}\left(f, R, U_{ext}\right) &\approx U_{ext} / \left(\dot{W}_{rest} / f + W\left(f, R, U_{ext}\right)\right), \\
\upsilon_{net}\left(f, R, U_{ext}\right) &\approx U_{ext} / W\left(f, R, U_{ext}\right), \\
\upsilon_{work}\left(f, R, U_{ext}\right) &\approx \left(\eta_h + \left(\varepsilon_h / 8\pi^2\right) U_{ext} / R^2 f\right)^{-1}, \\
\upsilon_{delta}\left(f, R, U_{ext}\right) &\approx \left(\eta_h + \left(\varepsilon_h / 4\pi^2\right) U_{ext} / R^2 f\right)^{-1}.
\end{aligned} \tag{17}$$

We illustrate the behavior of these measures in Fig 4 by providing estimates for typical efficiencies for Atzler et al.'s subject using the metabolic energy model in (8) with radius $R = 0.28$ m and avg. frequency $f = 0.3$ s$^{-1}$ for all external forces.

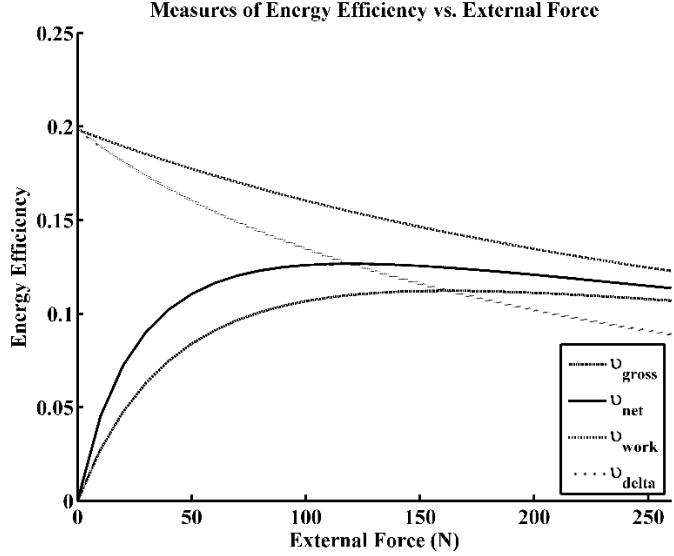

**Figure 4.** Measures of energy efficiency vs. external force. We give estimated functional forms of the gross, net, work, and delta energy efficiencies as the external force changes for a fixed case of asynchronous arm-cranking with radius $R = 0.28$ m and avg. frequency $f = 0.3$ s$^{-1}$.

We examine data from Powers et al. [17] where the energy efficiency are measured as functions of the external mechanical work rate $\dot{U}_{ext} = U_{ext} f$. The metabolic energy $W(f, R, \dot{U}_{ext})$ per turn as a function of the external mechanical work rate $\dot{U}_{ext}$ is:

$$W(f, R, \dot{U}_{ext}) = W_0 + \left(32\pi^4 \varepsilon_h m^2\right) R^2 f^3 \\ + \left(\varepsilon_h / 8\pi^2\right) \dot{U}_{ext}^2 / R^2 f^3 + \left(\eta_h\right) \dot{U}_{ext} / f. \tag{18}$$

The efficiencies become:

$$\begin{aligned} v_{gross}(f, R, \dot{U}_{ext}) &\approx \dot{U}_{ext} / \left(\dot{W}_{rest} + W(f, R, \dot{U}_{ext}) f\right), \\ v_{net}(f, R, \dot{U}_{ext}) &\approx \dot{U}_{ext} / W(f, R, \dot{U}_{ext}) f, \\ v_{work}(f, R, \dot{U}_{ext}) &\approx \left(\eta_h + \left(\varepsilon_h / 8\pi^2\right) \dot{U}_{ext} / R^2 f^2\right)^{-1}, \\ v_{delta}(f, R, \dot{U}_{ext}) &\approx \left(\eta_h + \left(\varepsilon_h / 4\pi^2\right) \dot{U}_{ext} / R^2 f^2\right)^{-1}. \end{aligned} \tag{19}$$

Recalling that 1.0 cal = 4.2 J, we observe that maximum work and delta work energy efficiencies are given by:

$$\max\{v_{work}\} \approx \max\{v_{delta}\} \approx 1 / \eta_h \approx 0.20. \tag{20}$$

We can compare the values estimated for the work, and delta work efficiencies in Fig. 4 to values measured empirically by Powers et al. [17] for frequencies of $f = 50$, 70, and 90 rpm (0.83, 1.2, and 1.5 s$^{-1}$) at external mechanical work rates of $\dot{U}_{ext} = 15$, 30, 45, and 60 W. We note that the frequencies observed by Powers et al. are all much higher than the $f = 18$ rpm (0.30 s$^{-1}$) observed by Atzler et al. [16], while the external mechanical work rates are comparable to the $\dot{U}_{ext} = 19$ to 96 W observed by Atzler et al. Powers et al. estimated their subjects to have gross energy efficiencies rising from 0.06 to 0.15 as the mechanical work rate increased, work energy efficiencies falling from 0.29 to 0.20 as the mechanical work rate increased, and delta work efficiencies falling from 0.30 to 0.14 as the mechanical work rate increased over the same range. These observations agree qualitatively with those illustrated in Fig. 4. However we found in (20) that the maximum values of the work and delta work energy efficiencies estimated for Atzler et al.'s subject in (15) are 0.20 and therefore lower than the highest corresponding values observed by Powers et al.; we discuss this further in Sec. 4.4. It appears the estimate in (15) for $\eta_h$ is too high, and a value closer to those estimated in [5, 6] would provide results closer to those observed by Powers et al.; we discuss this further in Sec. 4.4. The results also agree qualitatively with those observed by Gaesser & Brooks [15] for subjects pedaling (i.e. asynchronous leg-cranking) an electrically braked cycle ergometer during steady-rate exercise albeit again with lower energy efficiency values.

*4.2 Maximizing Net Energy Efficiency Given a Fixed Avg. Frequency*

We look at the selection of asynchronous arm-cranking doing external mechanical work in which the subject's only goal is to maximize the net energy efficiency. We suppose the subject must turn the asynchronous arm-crank of radius $R$ a total of $n$ revolutions against an external force $F_{ext}$. We require the subject to turn the crank with a fixed avg. frequency $f^*$. The total metabolic energy expended is $W_{total}(f^*, R, F_{ext}) = nW(f^*, R, F_{ext})$. We define the metabolic energy per unit distance travelled by each hand as $\Phi(f^*, R, F_{ext}) = W_{total}(f^*, R, F_{ext})/2\pi nR$; this gives:

$$\Phi\left(f^*, R, F_{ext}\right) = W\left(f^*, R, F_{ext}\right) / 2\pi R. \tag{21}$$

Looking at the definition of the net energy efficiency $\upsilon_{net}(v, s, F_{ext})$ in (3), we find that minimizing the total metabolic energy expended is equivalent to maximizing the net energy efficiency.

The subject selects the case of asynchronous arm-cranking that minimizes $\Phi(f^*, R, F_{ext})$ given an external force $F_{ext}$; this minimum is the solution to the equation:

$$\frac{\partial \Phi}{\partial R}\left(f^*, R, F_{cart}\right) = 0. \tag{22}$$

Combining (8) and (22), we find:

$$R^*\left(f^*, F_{ext}\right) \approx \frac{1}{8\pi^2 m \cdot \left(f^*\right)^2} \sqrt{\frac{2W_0 f^*}{\varepsilon_h} + F_{ext}^2}. \tag{23}$$

We recall that in Sec. 3.5 we noted that the estimate for the parameters $W_0$ was not statistically significant and that therefore its value could be very small or even zero. For sufficiently large $F_{ext}$, we find:

$$R^*\left(f^*, F_{ext}\right) \approx F_{ext} \Big/ 8\pi^2 m \cdot \left(f^*\right)^2. \tag{24}$$

Thus, for sufficiently large $F_{ext}$ the radius $R^*$ in (24) that maximizes the net energy efficiency is independent of the estimated values for the parameters $W_0$, $\varepsilon_h$, and $\eta_h$. However, there is clearly a limit on how long the radius $R^*$ may be that arises from the biomechanics of the arm.

*4.3 Predicted Movement Patterns*

We illustrate the maximum net energy efficiency asynchronous arm-cranking given a fixed avg. frequency using the case in which a subject selects an avg. frequency $f^* = 0.30$ s$^{-1}$ with a range of external forces $F_{ext} = 28$ N to 260 N as was used by Atzler et al.'s subject in [16] and Sec 3.5. Using (24), we estimate:

$$\begin{aligned} R^*\left(f^* = 0.30 \text{ s}^{-1}, F_{ext} = 28N\right) &\approx 1.3 \ m, \\ R^*\left(f^* = 0.30 \text{ s}^{-1}, F_{ext} = 260N\right) &\approx 12 \ m. \end{aligned} \quad (25)$$

Assuming the model in (8) holds at higher frequencies of rotation (however, see Sec. 4.4), we can estimate the optimal radii at higher frequencies. We use the highest and lowest frequencies $f^* = 0.83$ s$^{-1}$ and 1.5 s$^{-1}$ observed by Powers et al. [17] For the lowest frequency of $f^* = 0.83$ s$^{-1}$, we estimate:

$$\begin{aligned} R^*\left(f^* = 0.83 \text{ s}^{-1}, F_{ext} = 28N\right) &\approx 0.17 \ m, \\ R^*\left(f^* = 0.83 \text{ s}^{-1}, F_{ext} = 260N\right) &\approx 1.6 \ m. \end{aligned} \quad (26)$$

Using the highest frequency of $f^* = 1.5$ s$^{-1}$, we estimate:

$$\begin{aligned} R^*\left(f^* = 1.5 \text{ s}^{-1}, F_{ext} = 28N\right) &\approx 0.05 \ m, \\ R^*\left(f^* = 1.5 \text{ s}^{-1}, F_{ext} = 260N\right) &\approx 0.49 \ m. \end{aligned} \quad (27)$$

There is a limit on how long the radius $R^*$ may be that arises the biomechanics of the arm; we discuss these estimates further in Sec. 4.4.

*4.4 Discussion*

While the work and delta work energy efficiencies we have derived from the model developed in Sec. 3.3 agree qualitatively with Powers et al., [17] the energy efficiencies predicted by the model are somewhat lower. We have observed in (20) that the maximum values of the work and delta work energy efficiencies are determined by the parameter value $\eta_h$. We noted in Sec. 3.5 that the value for $\eta_h$ that we have estimated in (15) is roughly twice values estimated for comparable parameters in [4-6], and we have discussed in Sec. 3.6 a number of issues related to the model and the available data that might lead to a poor fit for $\eta_h$.

We have observed in (24) that when the external force $F_{ext}$ is sufficiently large the radius of the crank-handles of the asynchronous arm-crank giving maximum net energy efficiency is approximately independent of the parameter values estimated in (15), and have given example values for the radii $R^*$ of the crank-handles on the crank-arms giving maximum net efficiency in (25), (26), and (27) under various conditions. Many of the estimated radii $R^*$ are impractically long given the biomechanics of the arm, so it appears that, for larger external forces $F_{ext}$, the crank should be constructed with the largest radius $R^*$ that can be used comfortably by the subject; this would agree with what seems to be done in practice. However, the model for the optimal radius $R^*$ in (24) depends on the model for the force generated by each arm in (6) and (7), and in particular the centripetal force component $F_{cent}(f, R)$, and changes to the model of how the subject applies force to the crank may affected the estimated optimal radius $R^*$. We have discussed this modeling choice as well as other possible models for the force generated by the hands in Sec. 3.6.

## 5 Mechanical Advantage and Asynchronous Arm-Cranking

In the fourth part of this project, we modify the asynchronous arm-crank in Secs. 3 and 4 by introducing an adjustable mechanical advantage. For clarity and to facilitate comparison to a similar idealized mechanism driven by walking gait that we have analyzed in [6], we specify that the external work provides mechanical energy to a cart that is attached to a cord that, in turn, is attached to a spool that that is attached to the crank so that the rotation of the crank winds the cord winds the cord about the spool thus pulling the cart. We analyze the case where the subject performs asynchronous arm-cranking with fixed avg. frequency and calculate the mechanical advantage that maximizes the net energy efficiency.

*5.1 External Mechanical Work with Mechanical Advantage*

We imagine the subject is pulling a cart using an idealized mechanical device consisting of an asynchronous arm-crank attached to a spool so that the crank and spool rotate together around a common axle with the same frequency. Both the crank and spool are massless; the axle is frictionless. The crank-handles on the crank-arms are positioned at a radius $R$ from the center where $R$ has been fixed at some value that the subject finds comfortable. The spool has a radius $r$ that the subject is free to adjust by swapping out using spools of arbitrary radius. The spool attaches to a length of massless cord of infinitesimal thickness which can be wound around the spool. The cart is attached to the cord so that, as the spool rotates, the cord winds itself around a spool thereby pulling the cart. The cart requires a force $F_{cart}$ be applied to move at a constant avg. speed. The subject turns the crank at a steady state with avg. frequency $f^*$ causing the cart to move with an avg. speed $v_{cart}$. As we did in the asynchronous arm-cranking model developed in Sec. 3, we only look at steady state cranking that is in progress and maintains constant values for the movement parameters; we do not look at the process of starting or stopping cranking.

The torque $\tau$ required on the spool to pull the cart is $\tau = F_{cart} r$, so the subject must apply an external force satisfying $\tau = F_{ext} R$ to generate the required $F_{cart}$. In one rotation, the hands on the asynchronous arm-crank each travel a distance $2\pi R$ while the cart drawn by the cord attached to the spool travels a distance $2\pi r$. If one rotation takes a time $T$, the avg. speed of the hands is $v = 2\pi R/T$ and the avg. speed of the cart is $v_{cart} = 2\pi r/T$; we find:

$$v_{cart} / v = F_{ext} / F_{cart} = r / R. \tag{28}$$

The cost of adjusting the mechanical advantage $F_{ext}/F_{cart}$ so that the subject need only generate a lower external fore $F_{ext}$ is that the cart moves slower. In one rotation of the asynchronous arm-crank and spool the cart receives a mechanical energy of $U_{cart} = 2\pi F_{cart} r$, while the subject generates a mechanical energy of $U_{ext} = 2\pi F_{ext} R$, so we find $U_{cart} = U_{ext}$. Thus, all the generated mechanical energy above that which is a part of asynchronous arm-cranking when no external mechanical work is done is given to the cart as external mechanical work. Combining (10) and (28) gives a metabolic energy per turn of:

$$W\left(f^*, r, F_{cart}\right) \approx W_{turn}\left(f^*\right) + \left(\varepsilon_h / 2\right)\left(r / R\right)^2 F_{cart}^2 / f^* \qquad (29)$$
$$+ \left(2\pi\eta_h\right) F_{cart} r.$$

*5.2 Maximizing Net Energy Efficiency Given a Fixed Avg. Frequency*

We suppose that the subject must pull the cart a distance $D$. To do this, the subject must turn the spool $n = D/2\pi r$ times, and the total metabolic energy expended is $W_{total}(f^*, r, F_{ext}) = (D/2\pi r)W(f^*, r, F_{ext})$. We define the metabolic energy per unit distance $\Phi(f^*, r, F_{ext}) = W_{total}(f^*, r, F_{ext})/D$; this gives:

$$\Phi\left(f^*, r, F_{ext}\right) = W\left(f^*, r, F_{ext}\right) / 2\pi r. \qquad (30)$$

The asynchronous arm-cranking doing external work expending the minimum metabolic energy to move the cart solves the equation:

$$\frac{\partial \Phi}{\partial r}\left(f^*, r, F_{cart}\right) = 0. \qquad (31)$$

Combining (8), (29), and (31), and noting that the average metabolic power $\langle \dot{W}_{turn}(f) \rangle$ of cranking with no external load is given by $\langle \dot{W}_{turn}(f) \rangle = W_{turn}(f) \cdot f$, we find:

$$r^*\left(f^*, F_{cart}\right) / R = \sqrt{2\left\langle \dot{W}_{turn}\left(f^*\right)\right\rangle / \varepsilon_h} / F_{cart}. \qquad (32)$$

Combining (28) and (32) gives the external force:

$$F_{ext} = \sqrt{2\left\langle \dot{W}_{turn}\left(f^*\right)\right\rangle / \varepsilon_h}. \qquad (33)$$

Thus, given the avg. frequency $f^*$ and the radius $R$, if the mechanical advantage is chosen according to (32), then the external force the subject must generate is constant.

*5.3 Predicted Movement Patterns*

We illustrate the maximum net energy efficiency avg. frequency and mechanical advantage derived in Sec. 5.2 using the case in which a subject selects a radius $R = 0.36$ m. We first place the asynchronous arm-crank below shoulder height and use the parameter values in (15) with $\varepsilon_h(0.33 \leq \rho \leq 0.67)$; we estimate:

$$\begin{aligned} r^*\left(F_{cart}\right) / R &\approx \left[49\,N\right] / F_{cart}, \\ F_{ext} &\approx 49\,N, \\ v_{cart}\left(F_{cart}\right) &\approx \left[34\,N \cdot m \cdot s^{-1}\right] / F_{cart}. \end{aligned} \qquad (34)$$

For a cart pulled by Atzler et al.'s subject requiring forces $F_{cart}$ of 28 N and 260 N, the subject pulls the cart by generating an external force $F_{ext}$ of 49 N with the carts travelling at speeds $v_{cart}$ of 1.2 m·s⁻¹ and 0.13 m·s⁻¹, respectively.

We next place the asynchronous arm-crank above shoulder height use the parameter values in (15) with $\varepsilon_h(\rho = 0.95)$; we estimate:

$$\begin{aligned}
r^*\left(F_{cart}\right) / R &\approx \left[30\,N\right] / F_{cart}, \\
F_{ext} &\approx 30\,N, \\
v_{cart}\left(F_{cart}\right) &\approx \left[20\,N \cdot m \cdot s^{-1}\right] / F_{cart}.
\end{aligned} \qquad (35)$$

For a cart pulled by Atzler & Herbst's subject, the subject pulls the cart by generating an external force $F_{ext}$ of 30 N with the carts travelling at speeds $v_{cart}$ of 0.71 m · s$^{-1}$ and 0.077 m · s$^{-1}$, respectively.

*5.4 Discussion*

An alternative approach to avoiding the very slow movement that happens when the subject simply tries to maximize the net energy efficiency (Sec. 5.2) when fixing the avg. frequency $f$ (Sec. 5.3) would be to fix the avg. speed $v_{cart}$ of the cart, effectively fixing the time the task takes to complete. The solution of this problem would follow an approach analogous to the one in Sec. 5.3. Fixing the avg. frequency gives combinations of cases of asynchronous arm-cranking and mechanical advantage that fixes the external force the subject must generate though at the cost of possibly taking a long time to complete the task, while fixing the avg. speed $v_{cart}$ of the cart fixes the time taken to complete the task though at the cost of possibly using more onerous cases of cranking and generating larger external forces.

The idealized mechanical device of massless and frictionless asynchronous arm-cranks, spools, and cords is intended to provide clarity as to how mechanical advantage relates the task being performed to the external force that must be generated and the metabolic energies expended generating muscle forces and mechanical energy. We can replace the idealized mechanical device that we have used with a somewhat more practical, but still idealized one, one consisting of two spools of fixed radii each attached to a gear whose radius can be varied and with the two gears attached by a chain where the component parts are still massless and frictionless. This modified device begins to resemble the system of pedals (i.e. asynchronous leg-crank) and gears on bicycle and we see that the mechanical advantage becomes expressed in terms of the radius and gear ratio.